\documentclass[aps,prd,reprint,preprintnumbers,superscriptaddress,showpacs,twocolumn]{revtex4-1}
\usepackage{latexsym,graphicx,amssymb,amsmath,mathrsfs}
\usepackage{setspace,bm}
\usepackage[breaklinks, colorlinks=true, pdfstartview=FitV, linkcolor=red, citecolor=blue, urlcolor=blue]{hyperref}
\usepackage[usenames]{color}
\usepackage{latexsym}
\usepackage{epstopdf}
\usepackage{mathtools}
\usepackage{url}
\usepackage{comment}
\usepackage{braket}
\usepackage{CJKutf8}

\newcommand{\bequ}{\begin{equation}}
\newcommand{\eequ}{\end{equation}}
\newcommand{\bea}{\begin{eqnarray}}
\newcommand{\eea}{\end{eqnarray}}

\newcommand{\m}{\mu}

\DeclareSymbolFont{boldletters}{OML}{cmm} {b}{it}
\DeclareSymbolFontAlphabet{\mathbit}{boldletters}
\DeclareMathSymbol{\alpha}{\mathalpha}{letters}{"0B}
\DeclareMathSymbol{\beta}{\mathalpha}{letters}{"0C}
\DeclareMathSymbol{\gamma}{\mathalpha}{letters}{"0D}
\DeclareMathSymbol{\delta}{\mathalpha}{letters}{"0E}
\DeclareMathSymbol{\epsilon}{\mathalpha}{letters}{"0F}
\DeclareMathSymbol{\zeta}{\mathalpha}{letters}{"10}
\DeclareMathSymbol{\eta}{\mathalpha}{letters}{"11}
\DeclareMathSymbol{\theta}{\mathalpha}{letters}{"12}
\DeclareMathSymbol{\iota}{\mathalpha}{letters}{"13}
\DeclareMathSymbol{\kappa}{\mathalpha}{letters}{"14}
\DeclareMathSymbol{\lambda}{\mathalpha}{letters}{"15}
\DeclareMathSymbol{\mu}{\mathalpha}{letters}{"16}
\DeclareMathSymbol{\nu}{\mathalpha}{letters}{"17}
\DeclareMathSymbol{\xi}{\mathalpha}{letters}{"18}
\DeclareMathSymbol{\pi}{\mathalpha}{letters}{"19}
\DeclareMathSymbol{\rho}{\mathalpha}{letters}{"1A}
\DeclareMathSymbol{\sigma}{\mathalpha}{letters}{"1B}
\DeclareMathSymbol{\tau}{\mathalpha}{letters}{"1C}
\DeclareMathSymbol{\upsilon}{\mathalpha}{letters}{"1D}
\DeclareMathSymbol{\phi}{\mathalpha}{letters}{"1E}
\DeclareMathSymbol{\chi}{\mathalpha}{letters}{"1F}
\DeclareMathSymbol{\psi}{\mathalpha}{letters}{"20}
\DeclareMathSymbol{\omega}{\mathalpha}{letters}{"21}
\DeclareMathSymbol{\varepsilon}{\mathalpha}{letters}{"22}
\DeclareMathSymbol{\vartheta}{\mathalpha}{letters}{"23}
\DeclareMathSymbol{\varpi}{\mathalpha}{letters}{"24}
\DeclareMathSymbol{\varrho}{\mathalpha}{letters}{"25}
\DeclareMathSymbol{\varsigma}{\mathalpha}{letters}{"26}
\DeclareMathSymbol{\varphi}{\mathalpha}{letters}{"27}
\DeclareMathSymbol{\Gamma}{\mathalpha}{letters}{"00}
\DeclareMathSymbol{\Delta}{\mathalpha}{letters}{"01}
\DeclareMathSymbol{\Theta}{\mathalpha}{letters}{"02}
\DeclareMathSymbol{\Lambda}{\mathalpha}{letters}{"03}
\DeclareMathSymbol{\Xi}{\mathalpha}{letters}{"04}
\DeclareMathSymbol{\Pi}{\mathalpha}{letters}{"05}
\DeclareMathSymbol{\Sigma}{\mathalpha}{letters}{"06}
\DeclareMathSymbol{\Upsilon}{\mathalpha}{letters}{"07}
\DeclareMathSymbol{\Phi}{\mathalpha}{letters}{"08}
\DeclareMathSymbol{\Psi}{\mathalpha}{letters}{"09}
\DeclareMathSymbol{\Omega}{\mathalpha}{letters}{"0A}

\begin{document}
\title{Effective degrees of freedom, trace anomaly and c-theorem like condition in the hadron resonance gas model}

\author{Hiroaki Kouno}
\email[]{kounoh@cc.saga-u.ac.jp}
\affiliation{Department of Physics, Saga University,
             Saga 840-8502, Japan}

\author{Riki Oshima}
\email[]{24804001@edu.cc.saga-u.ac.jp}
\affiliation{Department of Physics, Saga University,
             Saga 840-8502, Japan}

\author{Kouji Kashiwa}
\email[]{kashiwa@fit.ac.jp}
\affiliation{Fukuoka Institute of Technology, Wajiro, Fukuoka 811-0295, Japan}


\begin{abstract}
The relation between the effective degrees of freedom (EDOF) and the trace anomaly is studied in the hadron resonance gas (HRG) model.  
If we regard the thermodynamical relation as the evolution equation and define the EDOF as $P/T^4$, where $P$ and $T$ are the pressure and the temperature, respectively, 
we obtain the equation which relates to the trace anomaly. 
The structure of the equation resembles that of the so-called c-theorem, which asserts that the EDOF should not increase as the energy scale parameter decreases, in the two dimensional conformal field theory. 
There is a stationary point where the trace anomaly (modified trace anomaly) vanishes, and the scale symmetry is restored. 
To investigate the limiting temperature of the HRG model with the excluded volume effects,  we consider two types of the c-theorem like conditions for the EDOF. 
The first condition requires that the EDOF should not decrease when $T$ increases. 
This condition is equivalent to the condition that the trace anomaly (modified trace anomaly) should not be negative. 
The second condition requires that the EDOF should be convex downwards as a function of $T$. 
It is found that the first condition gives the limiting temperature of the HRG model with the excluded volume effect which is much higher than the crossover transition temperature obtained by the lattice QCD calculation and, at zero baryon number density, is close to the transition temperature in the pure gluonic theory, while the second one gives the limiting temperature which almost coincides with the one obtained by using the normalized baryon number fluctuation in the previous study and is consistent with the critical point predicted by the lattice QCD calculation. 
\end{abstract}

\maketitle


\section{Introduction}
\label{intro}

The deconfinement of quarks is a mysterious phenomenon whose mechanism is not clear yet.  
Due to the strong interaction, at low temperature and low density, quarks are confined in hadrons, namely baryons and mesons. 
The lattice QCD (LQCD) calculations indicate that, at high temperature, quarks are deconfined and form quark matter. 
However, at $\mu =0$ where $\mu$ is the baryon number chemical potential, the LQCD calculations predict that the transition from hadron matter to quark matter    
is not a discontinuous phase transition but a continuous crossover transition~\cite{Aoki:2006we}.  
It is not well understood how quarks break free from hadrons during the crossover transition. 
Several new intermediate phases between the hadron phase and the ideal quark gluon plasma (QGP) phase are proposed; e.g., see Refs.~\cite{Pisarski:2006hz,Glozman:2014mka,Cohen:2023hbq,Fujimoto:2025sxx} and references therein.  

There is a fundamental question of at how high temperature hadrons can exist.
In a simple picture, as temperature ($T$) increases, the density of hadrons increases rapidly, and many hadrons combine and form quark matter.  
In this picture, it is essential that a hadron has a finite size, and it is expected that the limiting temperature of hadrons depends on their size. 

At low temperature, the equation of state obtained by LQCD calculations can be well reproduced by the hadron resonance gas (HRG) model. 
In the HRG model, the ideal gas approximation is usually used. 
In such an approximation, as temperature increases, the effective degrees of freedom (EDOF) increase rapidly, and the HRG results deviate from the LQCD ones. 
The ideal gas model has no mechanism that limits the model itself, since there is no interaction in the model, although the resonances themselves are produced by the interaction. 
To obtain the limiting temperature of hadrons, in particular, the repulsive interaction that makes hadrons unstable is needed. 
The repulsive interaction reduces the EDOF of hadron matter and makes hadron matter thermodynamically unfavorable.  
The more the effective degrees of freedom of hadron increase, the stronger the repulsive effects between hadrons become. 

One traditional way to incorporate the repulsive forces between hadrons is to take the excluded volume effect (EVE) into account~\cite{Cleymans:1986cq, Kouno:1988bi, Rischke:1991ke}; for the recent review, see, e.g., Ref.~\cite{Fujimoto:2021dvn} and references therein. 
In the HRG model, the EVE makes the baryon number density saturate at large $\mu$. 
The EVE was also investigated in the imaginary $\mu$ region where the LQCD is free from the infamous sign problem and is feasible; for the sign problem, see Refs.\,\cite{deForcrand:2010ys,Nagata:2021bru,*Nagata:2021ugx} as an example. 
It was found that the HRG model with EVE has a singularity at $\mu =i(2k+1)\pi T$ and $T\sim 0.2$ GeV where $k$ is any integer \,\cite{Taradiy:2019taz,Savchuk:2019yxl,Vovchenko:2017xad,Oshima:2023bip,Oshima:2026bub}.  
This singularity is a counterpart of the baryon density saturation at large real $\mu$ and corresponds to the Roberge-Weiss (RW) transition known in the QCD at imaginary $\mu$~\cite{Roberge:1986mm}. 
Hence, we call this singularity the Roberge-Weiss like (RWL) singularity in this paper. 
Obviously, the RWL temperature $T_{\rm RWL}\sim 0.2$ GeV is the limiting temperature of the HRG model with EVE at $\mu =i(2k+1)\pi T$. 
It should be remarked that the value of $T_{\rm RWL}$ depends on the degrees of freedom (DOF) of baryons and the baryon volume $v_{\rm B}$.  
The greater the degrees of freedom and/or the baryon volume, the lower the RWL temperature $T_{\rm RWL}$~\cite{Oshima:2023bip,Oshima:2026bub}. 
Is there any limiting temperature of the HRG model with EVE when $\mu$ is real? 
  
\begin{figure}[t]
\centering
\centerline{\includegraphics[width=0.40\textwidth]{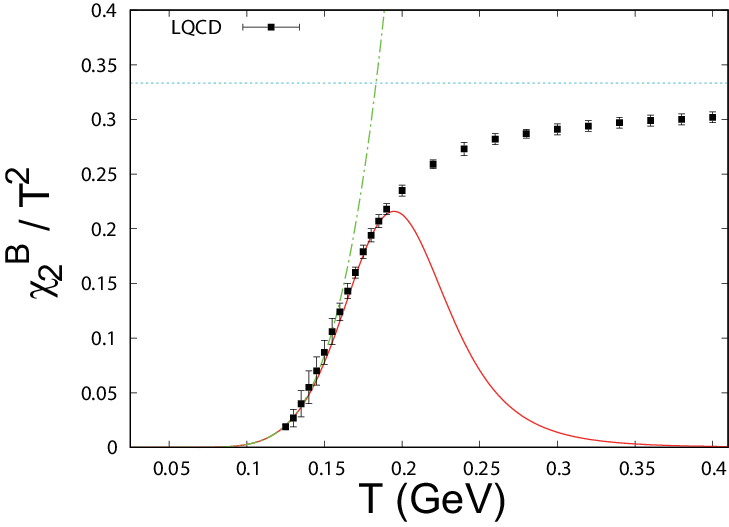}}
\caption{The solid (dash-dotted) line shows the $T$-dependence of the dimensionless baryon number fluctuation $\chi_2^{\rm B}/T^2$ at $\mu =0$ in the HRG model with EVE  (without EVE). 
For the detailed description of the model, see Sec.~\ref{HRGEVE}.  
The dotted line shows result in the ideal massless three flavor QGP.  
The squares with error bar show the LQCD results in Ref.~\cite{Borsanyi:2011sw}. 
 }
 \label{Fig_sus}
\end{figure}

In Ref.~\cite{Oshima:2026bub}, the baryon number fluctuation was used to investigate the limiting temperature of the baryon gas model. 
Figure~\ref{Fig_sus} shows $T$-dependence of the normalized baryon number fluctuation $\chi_2^{\rm B}/T^2$ at $\mu =0$ in the HRG models with EVE and without EVE; for details of the models, see Sec.~\ref{HRGEVE}. 
In the HRG model without EVE, $\chi_2^{\rm B}/T^2$ increases monotonically as $T$ increases. 
In the HRG model with EVE, it reaches its maximum at $T_{\chi, \rm max}=0.195$~GeV.  
We see that the result in the HRG model with EVE is consistent with that in LQCD up to $T_{\chi, \rm max}$. 
This indicates that $T_{\chi,\rm max}$ is the limiting temperature of the baryon gas model. 
In Ref.~\cite{Oshima:2026bub}, the $\mu$-dependence of $T_{\chi,\rm max}$ was also analyzed. 
It was found that the LQCD predicted critical point (CP) is located almost on the curve of $T_{\rm max}(\mu )$ in the $\mu$-$T$ plane. 
This limiting temperature is also related to the $R=0$ criterion, where $R$ is the scalar curvature in the thermodynamic geometry~\cite{Castorina:2018ayy,Castorina:2018gsx,Castorina:2019jzw,Zhang:2019neb,Castorina:2020vbh,Murgana:2023pyx}. 
However, the mechanism behind this coincidence is not clear yet.  

The dimensionless quantity $\chi_2^{\rm B}/T^2$ is related to the EDOF. 
In two dimensional conformal field theory, using the renormalization group method and the entropic theory, it was shown that the EDOF does not increase as the energy scale of the system decreases~\cite{Zamolodchikov:1986gt,Casini:2006es}. 
It is called the c-theorem. 
In the c-theorem, the c-function $c(r)$,  which represents the EDOF of the theory and converges to the central charge at the conformal fixed point, is defined as a function of the length scale parameter $r$ of the system, and $\frac{\partial c(r)}{\partial r}\leq 0$ is shown. 
This inequality ensures that the EDOF does not increase as $r$ increases.  
Inversely, it may be natural that the effective degrees of freedom do not decrease as the energy scale such as temperature increases. 
In this paper, to investigate the limiting temperature of the HRG model, we assume this condition for the normalized pressure which is defined as EDOF in our picture.   

\begin{figure}[t]
\centering
\centerline{\includegraphics[width=0.40\textwidth]{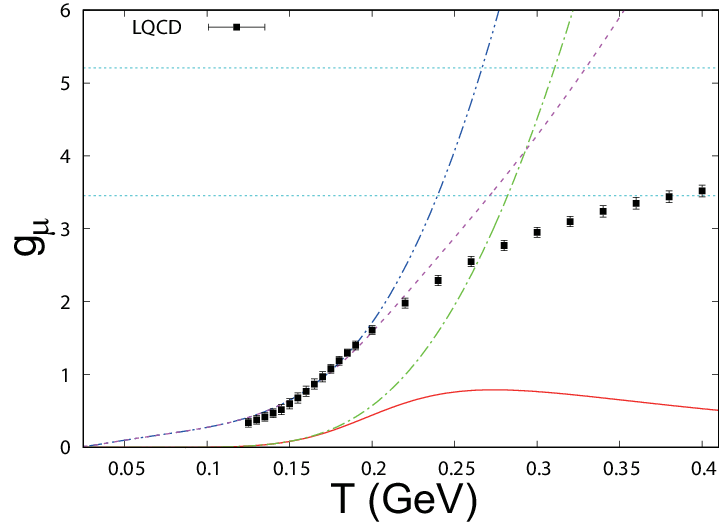}}
\caption{The $T$-dependence of the effective degrees of freedom $g_{\mu}(T)=P/T^4$ at $\mu =0$.  
The squares with error bar show the LQCD results in Ref.~\cite{Borsanyi:2012cr}. 
The upper (lower) dotted line shows result in the ideal massless three flavor QGP with (without) the gluon contributions. 
The dashed and dash-dot-dotted lines show the results in the HRG model with and without EVE, respectively. 
The solid, and dash-dotted lines show the baryon contribution $g_{\mu, B}$ in the HRG model with and without EVE, respectively. 
See Sec.~\ref{HRGEVE} for the detailed description of the HRG model and see Sec.~\ref{HRGEDOF} for the detail explanation of the lines. 
}
\label{Fig_cmu_mu=0}
\end{figure}

\begin{figure}[t]
\centering
\centerline{\includegraphics[width=0.40\textwidth]{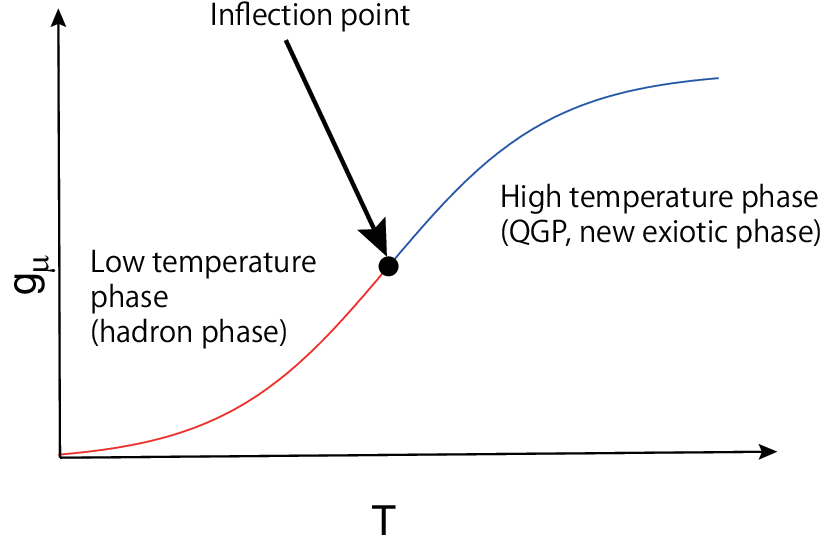}}
\caption{The solid line shows Schematic diagram of the effective degree of freedom in the transition from the low temperature phase to the high temperature phase. 
The solid circle shows the inflection point. 
}
\label{Fig_schematic}
\end{figure}

We also consider a stronger condition.  
Figure \ref{Fig_cmu_mu=0} shows the pressure $P$ divided by $T^4$, namely, the EDOF $g_\mu (T)$ in our picture, obtained by the LQCD calculation when $\mu =0$. 
The EDOF has an inflection point as a function of $T$ when $T\sim 0.175$~GeV. 
At the inflection point, $T^5$-normalized trace anomaly has its maximum. 
The EDOF is convex downwards above the inflection point and is convex upwards below it. 
In our simple picture, the region below the inflection point is regarded as the hadron phase, and the region above the point is regarded as the high temperature phase such as quark gluon plasma (QGP) or the exotic new phase; see Fig.~\ref{Fig_schematic}. 
(To say more precisely, since the transition is crossover at $\mu =0$, the low (high) temperature phase is composed mainly of hadron (QGP/exotic new matter). ) 
Hence, the EDOF of the hadron phase is convex downwards. 
We assume this condition for the EDOF in the HRG model. 

In this paper, to investigate the limitation of the hadron resonance gas model with the excluded volume effects, we study the effective degrees of freedom in the HRG model. 
This paper is organized as follows. 
In Sec.~\ref{EDofTA}, the relation between the EDOF and the trace anomaly is investigated. 
The relation between EDOF and the c-theorem is also discussed.  
In Sec.~\ref{HRGEVE}, we briefly show our formulation of the HRG model with EVE.  
In Sec.~\ref{HRGEDOF}, the EDOF is calculated in the HRG model with/without EVE . 
The limiting temperature in the HRG model with EVE is investigated.  
In Sec. \ref{EMESON}, the EVE in the meson gas is investigated. 
Section \ref{summary} is devoted to the summary and discussions.

\section{Effective degrees of freedom and the trace anomaly}
\label{EDofTA}

We start with the following thermodynamical relations
\begin{eqnarray}
{\partial P(T,\mu)\over{\partial T}}&=&s, 
\label{RG_T}
\\
{\partial P(T,\mu)\over{\partial \mu}}&=&n_{\rm B}, 
\label{RG_mu}
\end{eqnarray}
where $P$, $s$ and $n$ are the pressure, the entropy density, and the net baryon number density, respectively. 
We regard these equations as the evolution equations. 

We define the effective degrees of freedom $g_\mu (T) $ as 
\begin{eqnarray}
g_\mu (T) ={P(T,\mu )\over{T^4}}.
\label{effg_T}
\end{eqnarray}
Using the thermodynamic relation (\ref{RG_T}) with the aid of the thermodynamic identity $\varepsilon +P=Ts+\mu n_{\rm B}$,  we obtain 
\begin{eqnarray}
{\partial g_\mu (T) \over{\partial T}}={\varepsilon-3P-\mu n_{\rm B}\over{T^5}}={\Delta -\mu n_{\rm B}\over{T^5}}, 
\label{effg_c_T}
\end{eqnarray}
where $\varepsilon$ and $\Delta$ are the energy density and the trace anomaly, respectively.  
Hence, $g_\mu (T )$ increases (decreases) as $T$ increases if the modified trace anomaly $\Delta -\mu n_{\rm B}$ is positive (negative). 
${\partial g_\mu\over{\partial T}}$ vanishes  when $\Delta -\mu n_{\rm B}$ vanishes. 

Similarly, we can define the effective degrees of freedom $g_T(\mu )$ as 
\begin{eqnarray}
g_T (\mu) ={P(T, \mu )\over{\mu^4}}. 
\label{effg_mu}
\end{eqnarray}
Using the thermodynamic relation (\ref{RG_mu}), we obtain 
\begin{eqnarray}
{\partial g_T(\mu ) \over{\partial \mu }}={\epsilon -3P-Ts\over{\mu^5}} ={\Delta -Ts\over{\mu^5}}. 
\label{effg_c_mu}
\end{eqnarray}
Hence, 
$g_T(\mu )$ increases (decreases) as $\mu$ increases if the modified trace anomaly $\Delta -Ts$ is positive (negative).  
${\partial g_T\over{\partial \mu}}$ vanishes  when $\Delta -Ts$ vanishes. 

It may be more elegant to use the variable $(\beta =1/T,\gamma =-\mu/T)$ instead of  $(T,\mu)$. 
Note that $\beta$ has a dimension, but $\gamma$ does not. 
Since $\gamma$ depends not only on $\mu$ but also on $T$, the relation between the two coordinates $(T,\mu)$ and $(\beta,\gamma )$ is not simple, and we obtain
\begin{eqnarray}
{\partial\over{\partial\beta}}&=&-T^2{\partial\over{\partial T}}-T\mu{\partial\over{\partial \mu}},~~~~~
{\partial\over{\partial\gamma}}=-T{\partial\over{\partial \mu}}. 
\label{coordinate}
\end{eqnarray}
We define the effective degrees of freedom $g_\gamma (\beta )$ as 
\begin{eqnarray}
g_\gamma (\beta)  =\beta^4P(\beta, \gamma ). 
\label{effg}
\end{eqnarray}
Using the thermodynamic relations  (\ref{RG_T}) and (\ref{RG_mu}), we obtain 
\begin{eqnarray}
{\partial g_\gamma (\beta )\over{\partial \beta}}=-\beta^3(\varepsilon -3P)=-\beta^3\Delta. 
\label{effg_c_beta}
\end{eqnarray}
Hence, 
$g_\beta$ decreases (increases) as $\beta$ increases for the fixed value of $\gamma$, if the trace anomaly $\Delta =\varepsilon -3P$ is positive (negative). 
${\partial g_\gamma\over{\partial \beta}}$ vanishes when $\Delta$ vanishes, and the scale invariance is restored. 
This equation resembles the equation in the c-theorem~\cite{Zamolodchikov:1986gt,Casini:2006es} in the two dimensional conformal field theory. 
When $\Delta \ge 0$, the effective degrees of freedom do not increase as $\beta$ increases and converge to a constant value if there is a fixed point where the scale invariance is restored and ${\partial g_\gamma\over{\partial \beta}}=0$. 

Inversely, in the low energy effective theory, if we require that the EDOF of the effective theory should not decrease as the energy scale increases,  
we obtain the following conditions;
\begin{eqnarray}
{\partial g_\mu (T) \over{\partial T}}&=&{\Delta -\mu n_{\rm B}\over{T^5}}\ge 0, 
\label{effg_c_T_1}
\\
{\partial g_T(\mu ) \over{\partial \mu }}&=&{\Delta -Ts\over{\mu^5}}\ge 0, 
\label{effg_c_mu_1}
\\
{\partial g_\gamma (\beta )\over{\partial \beta}}&=&-\beta^3\Delta \le 0. 
\label{effg_c_beta_1}
\end{eqnarray}
These conditions may give the limiting energy scale of the low energy effective theory.  
We call these conditions the c-theorem like conditions in this paper. 

We also consider the stronger conditions;
\begin{eqnarray}
{\partial^2 g_\mu (T) \over{\partial T^2}}&=&{\partial\over{\partial T}}{\Delta -\mu n_{\rm B}\over{T^5}}\ge 0, 
\label{effg_c_T_2}
\\
{\partial^2 g_T(\mu ) \over{\partial \mu^2 }}&=&{\partial\over{\partial \mu}}{\Delta -Ts\over{\mu^4}}\ge 0, 
\label{effg_c_mu_2}
\\
{\partial^2 g_\gamma (\beta )\over{\partial \beta^2}}&=&-{\partial\over{\partial \beta}}\beta^3\Delta \ge 0. 
\label{effg_c_beta_2}
\end{eqnarray}
These conditions are equivalent to requiring that the EDOF be convex downward as a function of the scale parameter. 
If the right hand side of Eq. (\ref{effg_c_T_2}) (Eq. (\ref{effg_c_mu_2}), Eq. (\ref{effg_c_beta_2})) is zero and the condition (\ref{effg_c_T_1})  ((\ref{effg_c_mu_1}),  (\ref{effg_c_beta_1})) is satisfied, $(\Delta -\mu n_{\rm B})/T^5$ ($(\Delta -Ts)/\mu^5$, $\beta^3\Delta$) has its maximum there for the fixed value of $\mu$ ($T$, $\gamma$). 

It should be remarked that QCD itself does not satisfy the conditions (\ref{effg_c_T_2})$\sim$(\ref{effg_c_beta_2}). 
As will be seen in Sec.~\ref{HRGEDOF}, the LQCD calculation shows that the trace anomaly decreases as $T$ increases in the high temperature region. In fact, as was already seen in Sec.~\ref{intro}, in QCD, the EDOF is convex downwards as a function of $T$ in the low temperature region and is convex upwards in the high temperature region when $\mu =0$.  
In our simple picture, the system is in the hadron phase when the EDOF is convex downward, whereas it is in the quark phase when the EDOF is convex upward. 
The inflection point is regarded as the transition point. 
Hence, the HRG model should satisfy the conditions (\ref{effg_c_T_2})$\sim$(\ref{effg_c_beta_2}). 

In this paper, since we concentrate our discussions in the region of temperature $T=0.25\sim 0.410$ GeV, we consider the conditions (\ref{effg_c_T_1}), (\ref{effg_c_beta_1}), (\ref{effg_c_T_2}), and (\ref{effg_c_beta_2}).  
We show that the strong condition (\ref{effg_c_T_2}) gives the limiting temperature which is consistent with the one obtained by using the normalized baryon number fluctuation~\cite{Oshima:2026bub}.

\section{Hadron resonance gas model with excluded volume effects }
\label{HRGEVE}

In this section, we briefly review our hadron resonance gas model with EVE~\cite{Kouno:2023ygw,Oshima:2023bip,Kouno:2024cgo,Oshima:2026bub}. 
For simplicity, we assume that all baryons and antibaryons have the same volume $v_{\rm B}$ 
and put $v_{\rm B}={4\pi\over{3}}r_{\rm B}^3$ with $r_{\rm B}=0.8$~fm. 
Hereafter, we use the subscript B (M) to denote that the thermodynamic quantity with it is the baryon (meson) contribution of the quantity. 
The net baryon number density $n_{\rm B}$ is given by 
\begin{eqnarray}
n_{\rm B}(T,\mu )&=&n_{\rm b}(T,\mu )-n_{\rm a}(T,\mu ); 
\label{EnB}
\\
n_{\rm b}(T,\mu )&=&{n_{\rm b0}(T,\mu )\over{1+v_{\rm B}n_{\rm b0}(T,\mu )}}, 
\label{Enb}
\\
n_{\rm a}(T,\mu )&=&{n_{\rm a0}(T,\mu )\over{1+v_{\rm B}n_{\rm a0}(T,\mu )}}
\nonumber\\
&=&{n_{\rm b0}(T,-\mu )\over{1+v_{\rm B}n_{\rm b0}(T,-\mu )}}, 
\label{Ena}
\end{eqnarray}
where $n_{\rm b}$ and $n_{\rm a}$ are the number density of baryons and antibaryons, respectively, and $n_{\rm b0}$ and $n_{\rm a0}$ are these quantities calculated using the point particle approximation. 
When $\mu \to \pm \infty$, 
\begin{eqnarray}
n_{\rm B}(T,\mu )\to \pm {1\over{v_{\rm B}}}. 
\label{EnB_B_close}
\end{eqnarray}
In high density, the net baryon number density saturates to a constant value.   

Using the thermodynamic relation (\ref{RG_mu}), the pressure $P_B(T,\mu )$ of baryons and antibaryons is given by  
\begin{eqnarray}
P_{\rm B}(T,\mu ) &=& P_{\rm b}(T,\mu )+P_{\rm a}(T,\mu ); 
\label{PB_EVE}
\\
P_{\rm b}(T,\mu ) &=& \int d\mu n_{\rm b}(T,\mu ), 
\label{Pb_EVE}
\\
P_{\rm a}(T,\mu ) &=& -\int d\mu n_{\rm a}(T,\mu ) 
\nonumber\\
&=&P_{\rm b}(T,-\mu ). 
\label{Pa_EVE}
\end{eqnarray}
We adopt the natural boundary conditions $P_{\rm b}\to 0~(\mu \to -\infty)$ and $P_{\rm a}\to 0~(\mu \to \infty)$.   

The total pressure $P(T,\mu)$ of the HRG is given by
\begin{eqnarray}
P(T,\mu )=P_{\rm B}(T,\mu)+P_{\rm M}(T), 
\label{P_total}
\end{eqnarray}
where $P_{\rm M}$ is the meson pressure.   
Here, we use an ideal Bose gas approximation for mesons.  
The other thermodynamic quantities are calculated by using the thermodynamic relations. 
The EVE for mesons will be investigated later in Sec.~\ref{EMESON}. 

For baryons, we use the Boltzmann distribution function which is a good approximation of the Fermi distribution function unless the quantum effects are not large.   
In this approximation, we can obtain simple semi analytical representations 
for thermodynamic quantities~\cite{Oshima:2026bub},  
\begin{eqnarray}
n_{\rm b}(T,\mu )&=&{B(T)e^{\mu /T}\over{1+v_{\rm B}B(T)e^{\m/T}}}, 
\label{Enb_BA}
\\
n_{\rm a}(T,\mu )&=&{B(T)e^{-\mu /T}\over{1+v_{\rm B}B(T)e^{-\mu /T}}}, 
\label{Ena_BA}
\\
P_{\rm b}(T,\mu )&=&{T\over{v_{\rm B}}}\log{[1+v_{\rm B}B(T)e^{\mu /T}]}, 
\label{Pb_EVE_Boltz}
\\
P_{\rm a}(T,\mu )&=&{T\over{v_{\rm B}}}\log{[1+v_{\rm B}B(T)e^{-\mu /T}]}; 
\label{Pa_EVE_Boltz}
\end{eqnarray}
where
\begin{eqnarray}
B(T)&=&\sum_i B_i(T), 
\label{BT}
\\
B_i(T)&=&{g_{{\rm B},i}\over{2\pi^2}}\int_0^\infty dp p^2e^{-\sqrt{p^2+m_{{\rm B},i}^2}/T}, 
 \label{BiT}
\end{eqnarray}
here $m_{{\rm B},i}$ and $g_{{\rm B},i}$ are the mass and the spin degeneracy of $i$-th baryons (antibaryons), respectively.  
In the limit of $v_{\rm B}\to 0$, we obtain the net baryon number density and the pressure in the HRG without EVE;
\begin{eqnarray}
n_{\rm b0}(T,\mu )&=&B(T)e^{\mu /T},
\label{Enb_BA0}
\\
n_{\rm a0}(T,\mu )&=&B(T)e^{-\mu /T}, 
\label{Ena_BA0}
\\
P_{\rm b0}(T,\mu )&=&TB(T)e^{\mu /T}, 
\label{Pb_EVE_Boltz0}
\\
P_{\rm a0}(T,\mu )&=&TB(T)e^{-\mu /T}. 
\label{Pa_EVE_Boltz0}
\end{eqnarray}

\section{Effective degrees of freedom and trace anomaly in Hadron resonance gas model}
\label{HRGEDOF}

In the case of the point-like baryon gas model (\ref{Enb_BA0})$\sim$(\ref{Pa_EVE_Boltz0}) with the Boltzmann distribution, it can be easily shown that 
\begin{eqnarray}
g_{\mu, \rm B} (T) &=& \frac{\chi_2^{\rm B}}{T^2},~~
g_{T,\rm B} (\mu )= \frac{T^2\chi_2^{\rm B}}{\mu^4},
\nonumber\\
g_{\gamma,\rm B} (\beta ) &=& \beta^2\chi_2^{\rm B}. 
\label{effg_c_0}
\end{eqnarray}
Hence, 
\begin{eqnarray}
{\partial g_{\mu,\rm B} (T) \over{\partial T}}={\varepsilon_{\rm B}-3P_{\rm B}-\mu n_{\rm B}\over{T^5}}={\Delta_{\rm B} -\mu n_{\rm B}\over{T^5}}=0, 
\label{effg_c_T_0}
\end{eqnarray}
when $\chi_2^{\rm B}/T^2$ has its maximum/minimum.  
Similarly, 
\begin{eqnarray}
{\partial g_{T,\rm B} (\mu ) \over{\partial \mu}}={\varepsilon_{\rm B}-3P_{\rm B}-Ts_{\rm B}\over{\mu^5}}={\Delta_{\rm B} -Ts_{\rm B}\over{\mu^5}}=0, 
\label{effg_c_mu_0}
\end{eqnarray}
when $T^2\chi_2^{\rm B}/\mu^4$ has its maximum/minimum, and 
\begin{eqnarray}
{\partial g_{\gamma,\rm B} (\beta) \over{\partial \beta }}=-\beta^3(\varepsilon_{\rm B}-3P_{\rm B})=-\beta^3\Delta_{\rm B}=0, 
\label{effg_c_beta_0}
\end{eqnarray}
when $\beta^2\chi_2^{\rm B}$ has its maximum/minimum. 
In the case of Eqs. (\ref{Enb_BA0})$\sim$(\ref{Pa_EVE_Boltz0}), the EDOF and $\chi_2^{\rm B}/T^2$ increase as $T$ increases and saturate to a constant value in the high temperature limit where $m_{{\rm B},i}/T\to 0$, if the maximum of the baryon mass $m_{{\rm B},i}$ is finite. 
Hence, the point-like ideal baryon gas model has no limitations for the condition  (\ref{effg_c_T_1}).  
However, the excluded volume effects change the situation. 
In the following, we show the numerical results using the hadron resonance data in Ref.~\cite{ParticleDataGroup:2024cfk}.

\begin{figure}[t]
\centering
\centerline{\includegraphics[width=0.40\textwidth]{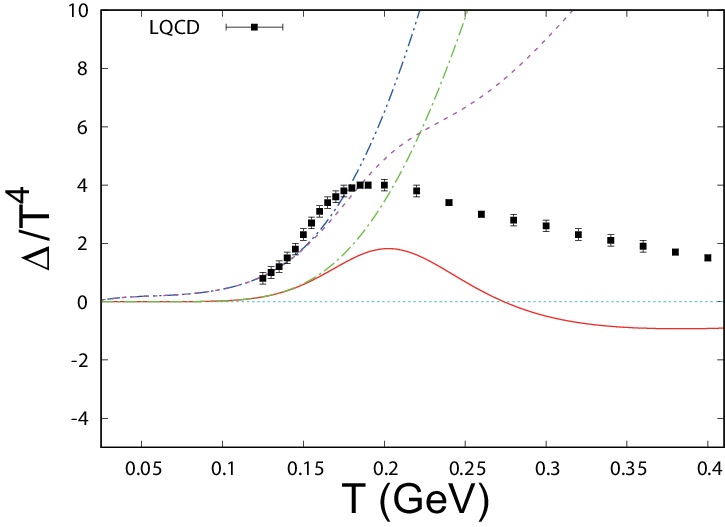}}
\caption{The $T$-dependence of the normalized trace anomaly $\Delta/T^4$ at $\mu =0$.  
The dashed and dash-dot-dotted lines show the results in the HRG model with and without EVE, respectively. 
The solid, and dash-dotted lines show the baryon contribution $\Delta_{\rm B}/T^5$ in the HRG model with and without EVE, respectively. 
The dotted line shows the result in the ideal massless three flavor QGP with (without) the gluon contributions. 
The squares with error bar show the LQCD results in Ref.~\cite{Borsanyi:2012cr}. 
 }
 \label{Fig_ta_mu=0}
\end{figure}

Figure~\ref{Fig_cmu_mu=0} in Sec.~\ref{intro} and Fig.~\ref{Fig_ta_mu=0} shows $g_\mu (T)=P/T^4$ and $\Delta /T^4$ as a function of $T$, respectively, in the HRG model with and without EVE, when $\mu =0$. 
Note that, at $\mu =0$, $g_\gamma (\beta ) =g_\mu (T)$ and $\Delta -\mu n_{\rm B}=\Delta$. 
In both cases,  $g_\mu (T)$ and $\Delta$ increase monotonically as $T$ increases. 
However, in the HRG model with EVE, the baryon contribution $g_{\mu,B}$ has its maximum at $T=0.274$ GeV where the baryon contribution $\Delta_B$ vanishes. 
The temperature $T=0.274$ GeV is much larger than the limiting temperature $T_{\chi,\rm max}=0.195$ GeV obtained by using $\chi_2^{\rm B}/T^2$, and is close to the transition temperature $T_{\rm d}=0.285$~GeV in pure gluonic theory~\cite{Borsanyi:2022xml,Giusti:2025fxu,Fujimoto:2025sxx}. 

On the other hand, $\Delta_{\rm B}$ has its maximum $T=0.202$ GeV.  
This temperature is close to $T_{\chi, \rm max}$.  
Note that the maximum point of $\Delta_{\rm B} /T^5$ ($\beta^3\Delta_{\rm B}$) corresponds to the inflection point of $g_{\mu,B} (T)$ ($g_{\gamma, B} (\beta )$).  
At this point, the right hand side of the second derivative (\ref{effg_c_T_2}) ((\ref{effg_c_beta_2})) vanishes.

\begin{figure}[t]
\centering
\centerline{\includegraphics[width=0.40\textwidth]{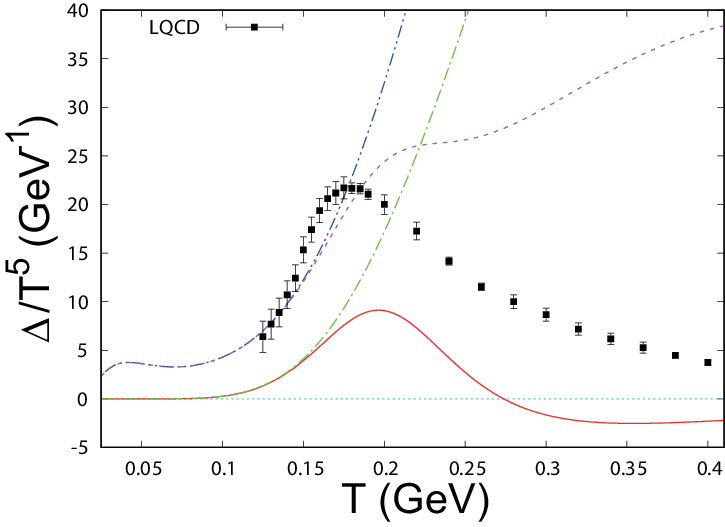}}
\caption{The $T$-dependence of the $T^5$-scaled trace anomaly $\Delta/T^5$ at $\mu =0$.  
The dashed and dash-dot-dotted lines show the results in the HRG model with and without EVE, respectively. 
The solid and dash-dotted lines show the baryon contribution $\Delta_{\rm B} /T^5$ in the HRG model with and without EVE, respectively.   
The dotted line shows result in the ideal massless three flavor QGP with (without) the gluon contributions. 
The squares with error bar show the LQCD results in Ref.~\cite{Borsanyi:2012cr}. 
 }
 \label{Fig_ta_5_mu=0_B_03}
\end{figure}

\begin{figure}[t]
\centering
\centerline{\includegraphics[width=0.40\textwidth]{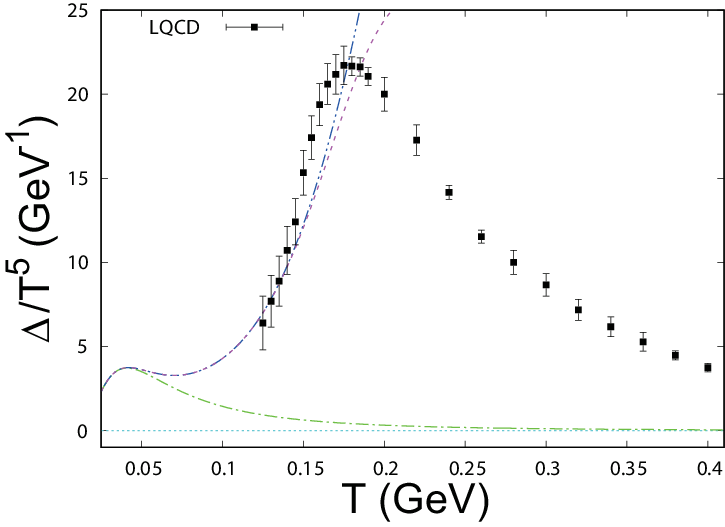}}
\caption{The $T$-dependence of the $T^5$-scaled trace anomaly $\Delta/T^5$ at $\mu =0$.  
The dashed and dash-dot-dotted lines show the results in the HRG model with and without EVE, respectively. 
The dash-dotted line shows the pion contribution $\Delta_{\pi} /T^5$ in the HRG model.   
The dotted line shows result in the ideal massless three flavor QGP with (without) the gluon contributions. 
The squares with error bar show the LQCD results in Ref.~\cite{Borsanyi:2012cr}. 
 }
 \label{Fig_ta_5_mu=0_pi_03}
\end{figure}

Fig.~\ref{Fig_ta_5_mu=0_B_03} shows the $T^5$-normalized trace anomaly $\Delta /T^5$ as a function of $T$ in the HRG model with and without EVE, when $\mu =0$. 
The baryon contribution $\Delta_{\rm B} /T^5$ increases monotonically as it increases in the HRG model without EVE, while it has its maximum at $T=0.197$ GeV in the HRG model with EVE.  
It should be also noted that $\Delta/T^5$ has its local maximum at $T=0.041$ GeV in the HRG model. 
This local maximum is induced by the pion contribution. 
Fig.~\ref{Fig_ta_5_mu=0_pi_03} shows the pion contribution to $\Delta/T^5$ in the HRG model.  
It is clear that this local peak is induced by the pion contribution. 
Since the pion is the pseudo Nambu-Goldstone boson and its mass is much smaller than those of the other hadrons, the pion contribution $\Delta_\pi/T^5$ has local peak in low temperature. This property is not unnatural, hence, we do not regard that this peak structure indicates the limitation of the HRG model.  
This peak structure remains in the model with EVE to mesons, since it locates in quite low temperature region. 
For the EVE to mesons, see Sec.~\ref{EMESON}.

\begin{figure}[t]
\centering
\centerline{\includegraphics[width=0.40\textwidth]{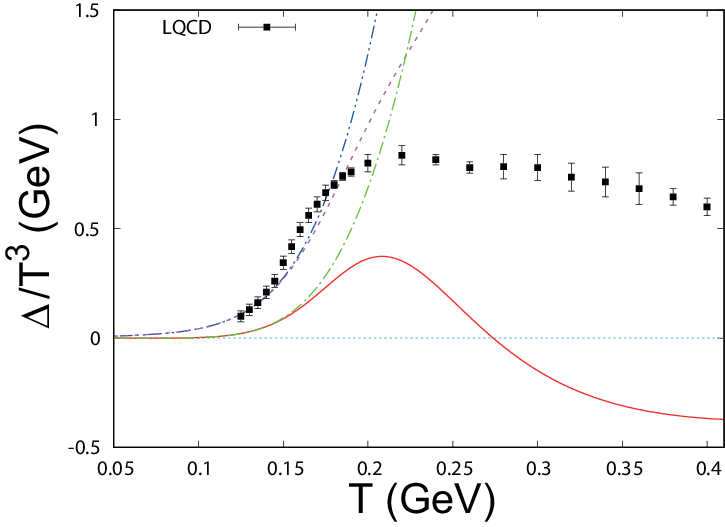}}
\caption{The $T$-dependence of the $\beta^3\Delta$ ($=\Delta/T^3$) at $\mu =0$.  
The dashed and dash-dot-dotted lines show the results in the HRG model with EVE and without EVE, respectively. 
The solid and dash-dotted lines show the baryon contribution $\beta^3\Delta_{\rm B} $ in the HRG model with and without EVE, respectively. 
The dotted line shows result in the ideal massless three flavor QGP with (without) the gluon contributions. 
The squares with error bar show the LQCD results in Ref.~\cite{Borsanyi:2012cr}. 
 }
 \label{Fig_ta_3_mu=0_B_beta} 
\end{figure}

We also show the result of $\beta^3\Delta$ in Fig.~\ref{Fig_ta_3_mu=0_B_beta}. 
In Ref.~\cite{Pisarski:2006hz}, it was pointed out that the $T^2$-normalized trace anomaly $\Delta/T^2$ is nearly constant in the pure gluonic theory, when $T=(1-4)T_{\rm d}$ where $T_{\rm d}$ is the transition temperature. 
In Fig.~\ref{Fig_ta_3_mu=0_B_beta}, it is seen that the $T^3$-normalized trace anomaly $\Delta/T^3$ in the LQCD calculations decreases very slowly as $T$ increases in the region of $T>0.22$ GeV, 
while it decreases rapidly in the HRG model with EVE.   

\begin{figure}[t]
\centering
\centerline{\includegraphics[width=0.40\textwidth]{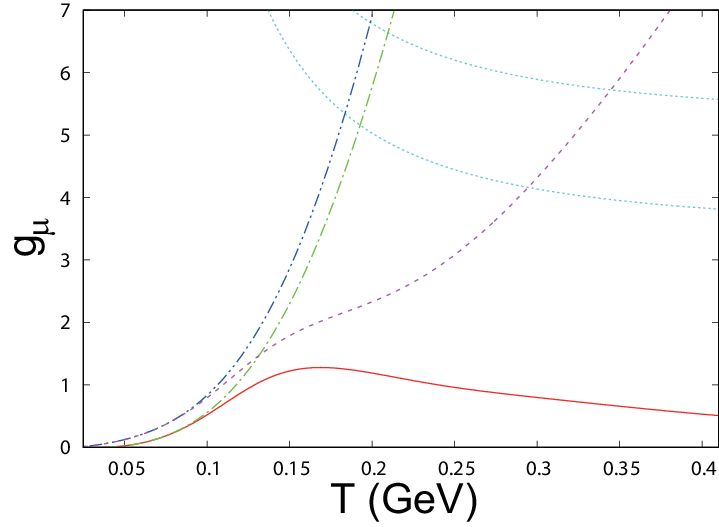}}
\caption{The $T$-dependence of the effective degrees of freedom $g_{\mu}$ at $\mu =0.6$ GeV.   
The dashed and dash-dot-dotted lines show the results in the HRG model with and without EVE, respectively. 
The solid and dash-dotted lines show the baryon contribution to $g_{\mu,{\rm B}}$ in the HRG model with and without EVE, respectively. 
The upper (lower) dotted line shows result in the ideal massless three flavor QGP with (without) the gluon contributions. 
 }
 \label{Fig_cmu_mu=600MeV}
\end{figure}

\begin{figure}[t]
\centering
\centerline{\includegraphics[width=0.40\textwidth]{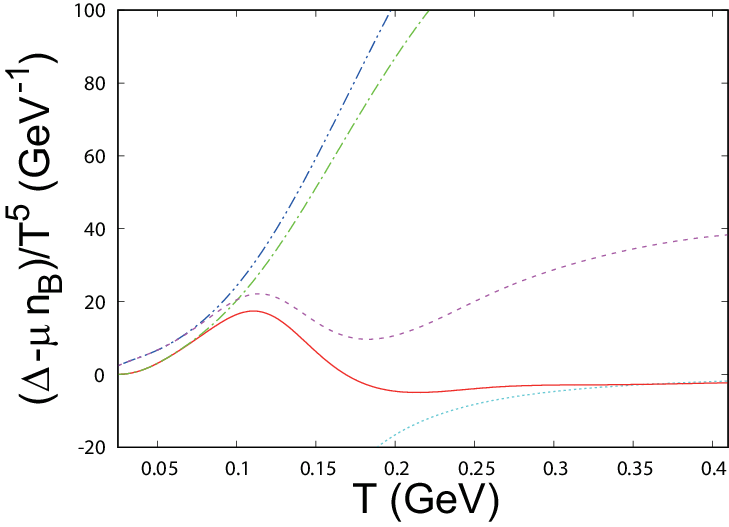}}
\caption{The $T$-dependence of $(\Delta -\mu n_{\rm B})/T^5$ at $\mu =0.6$ GeV.   
The dashed and dash-dot-dotted lines show the results in the HRG model with and without EVE, respectively. 
The solid and dash-dotted lines show the baryon contribution $(\Delta_{\rm B} -\mu n_{\rm B})/T^5$ in the HRG model with and without EVE, respectively. 
The dotted line shows the result in the ideal massless three flavor QGP with (without) the gluon contributions. 
 }
 \label{Fig_ta_mu=600MeV}
\end{figure}

Figure~\ref{Fig_cmu_mu=600MeV} and ~\ref{Fig_ta_mu=600MeV} show $g_\mu (T)$ and $(\Delta -\mu n_{\rm B})/T^5$ as a function of $T$, respectively, when $\mu =0.6$ GeV. 
In the HRG model without EVE, $g_\mu (T)$ and its baryon contribution $g_{\mu, \rm B}$ increase monotonically as $T$ increases. 
In the HRG model with EVE, $g_{\mu, B}$ has its maximum at $T=0.169$ GeV where $\Delta_{\rm B}-\mu n_{\rm B}$ vanishes. 
On the other hand, the maximum point of $(\Delta_{\rm B} -\mu n_{\rm B})/T^5$, namely $T=0.111$ GeV, corresponds to the inflection point of $g_{\mu, \rm B} (T)$.  
In the HRG model with EVE, the total $\Delta -\mu n_{\rm B}/T^5$ also has a local maximum at $T\sim 0.111$ GeV, since the baryon contribution is dominant in this case.  

\begin{figure}[t]
\centering
\centerline{\includegraphics[width=0.40\textwidth]{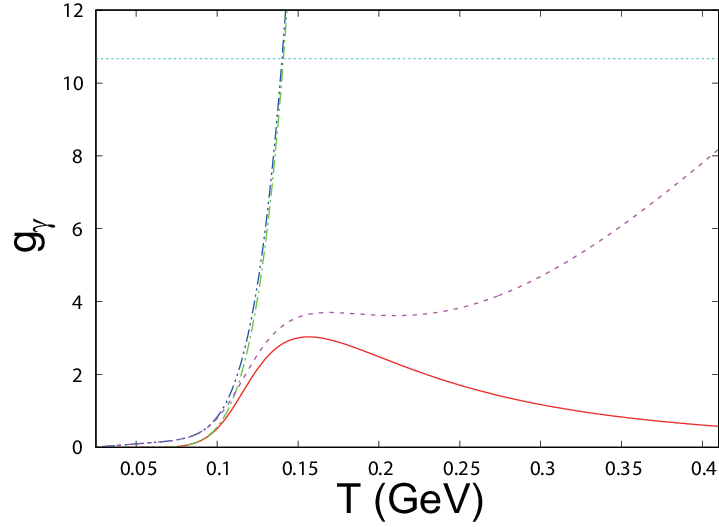}}
\caption{The $T$-dependence of the effective degrees of freedom $g_{\gamma}$ at $\gamma =-6$.  
The dashed and dash-dot-dotted lines show the results in the HRG model with and without EVE, respectively. 
The solid and dash-dotted lines show the baryon contribution $g_{\gamma,\rm B}$ in the HRG model with and without EVE, respectively. 
The upper (lower) dotted line shows result in the ideal massless three flavor QGP with (without) the gluon contributions. 
 }
 \label{Fig_cgamma_gamma=-6}
\end{figure}

\begin{figure}[t]
\centering
\centerline{\includegraphics[width=0.40\textwidth]{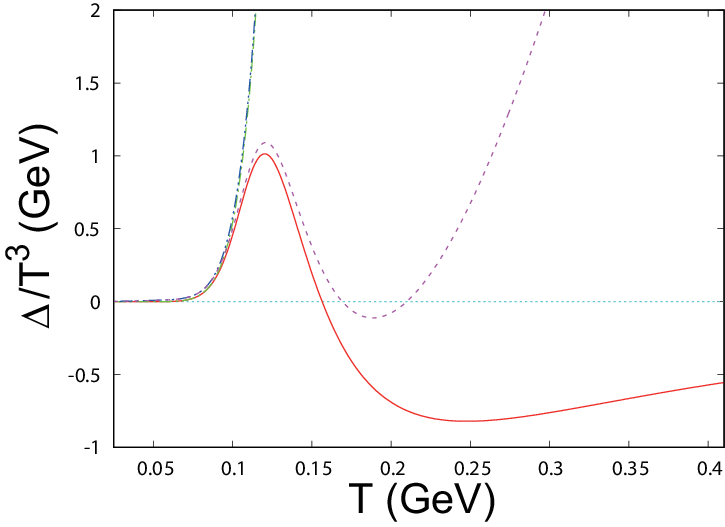}}
\caption{The $T$-dependence of the trace anomaly $\beta^3\Delta$ at $\gamma =-6$.  
The dashed, and dash-dot-dotted lines show the results in the HRG model with and without EVE, respectively. 
The solid, and dash-dotted lines show the baryon contribution $\beta^3\Delta_{\rm B}$ in the HRG model with and without EVE, respectively. 
The dotted line shows the result in the ideal massless three flavor QGP with (without) the gluon contributions. 
 }
 \label{Fig_ta_gamma=-6}
\end{figure}

Figure~\ref{Fig_cgamma_gamma=-6} and ~\ref{Fig_ta_gamma=-6} show $g_\gamma (\beta )$ and $\Delta /T^3$ as a function of $T$, respectively, when $\gamma =-6$. 
Since the baryon contribution is dominant at $\gamma =-\mu/T =-6$, the dashed-dotted and the dash-dot-dotted lines almost coincide in the HRG model without EVE. In the HRG model with EVE, the difference between the solid and dashed lines is not small, since the baryon contribution is suppressed by the excluded volume effects. 
In the case of the HRG model without EVE, $g_\gamma (\beta )$ increases monotonically as $T$ increases, while it has its local maximum (minimum) at  $T=0.169$ ($T=0.210$ GeV) where $\Delta$ vanishes, and its baryon contribution $g_{\gamma,\rm B} (\beta )$ has its maximum at $T=0.157$ GeV where $\Delta_{\rm B}$ vanishes in the HRG model with EVE, and the maximum point of $\beta^3\Delta_{\rm B}$, $T=1/\beta =0.120$ GeV, corresponds to the inflection point of $g_{\gamma,\rm B} (\beta )$.   
Since the baryon contribution is a main contribution at large $\gamma$,  $\beta^3\Delta$ also have their maxima at $T\sim 0.120$ GeV.

\begin{figure}[t]
\centering
\centerline{\includegraphics[width=0.40\textwidth]{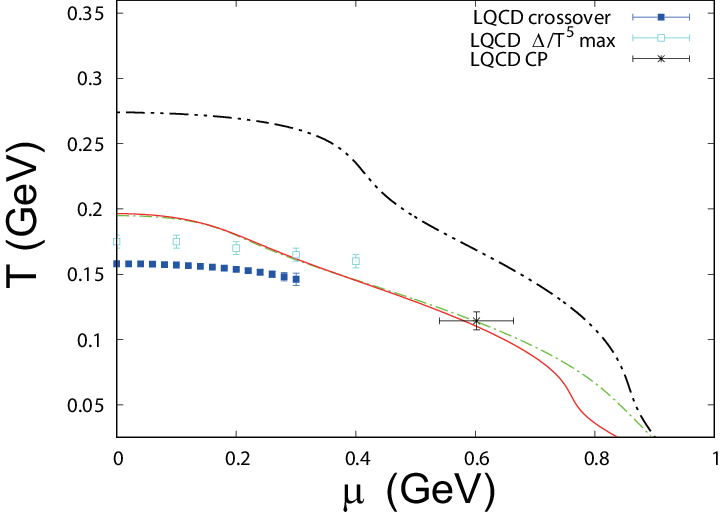}}
\caption{The dash-dot-dotted line shows the $T_{\rm zero}$ where $\Delta_{\rm B}-\mu n_{\rm B}$ vanishes in the HRG model with EVE. 
The solid (dash-dotted) line shows the temperature $T_{\rm max}$ ($T_{\chi ,\rm max}$) where $(\Delta_{\rm B}-\mu n_{\rm B})/T^5$ ($\chi_{2}^{\rm B}/T^2$) has its maximum for the fixed value of $\mu$.  
The solid squares with the error bar show the LQCD crossover line taken from Ref. ~\cite{Borsanyi:2020fev}. 
The open squares with the error bar show the temperature $T_{\rm max,L}$ where $\Delta/T^5$ has its maximum obtained by using the LQCD data in Ref. ~\cite{Borsanyi:2012cr}.  
The crosses with error bar shows the LQCD predicted critical point taken from Ref.~\cite{Shah:2024img}. 
 }
 \label{Fig_max_min}
\end{figure}

Figure~\ref{Fig_max_min} shows the $\mu$-dependence of the temperature $T_{\rm zero}$ where the baryon contribution of the modified trace anomaly $(\Delta_{\rm B}-\mu n_{\rm B})$ vanishes in the HRG model with EVE. 
The $T_{\rm zero}$ is the limiting temperature of the HRG model with EVE obtained by the c-theorem like condition (\ref{effg_c_T_1}), 
 and $(\Delta_{\rm B}-\mu n_{\rm B})/T^5$ is negative beyond $T_{\rm zero}$. 
At small $\mu$, the curve of $T_{\rm zero}$ is much higher than the LQCD crossover line~\cite{Borsanyi:2020fev} and, at $\mu =0$, is close to the transition temperature $T_{\rm d}\sim 0.285$ GeV in the pure gluonic theory~\cite{Borsanyi:2022xml,Giusti:2025fxu,Fujimoto:2025sxx}. 
At $\mu =\mu_{\rm CP}$, $T_{\rm zero}$ is much higher than the temperature $T_{\rm CP}$, where $\mu_{\rm CP}$ and $T_{\rm CP}$ are the baryon number chemical potential and the temperature at the critical point predicted by the LQCD~\cite{Shah:2024img}, respectively.  

In Fig.~\ref{Fig_max_min}, 
the curves of $T_{\rm max }(\mu )$ and $T_{\chi, \rm max}(\mu )$ where $(\Delta_{\rm B} -\mu n_{\rm B})/T^5$ and  $\chi_2^{\rm B}/T^2$ have their maximum for the fixed value of $\mu$ in the HRG model with EVE, respectively, are also shown. 
The $T_{\rm max}$ is the limiting temperature of the HRG model with EVE obtained by the strong condition (\ref{effg_c_T_2}).  
We see that the critical point predicted by the LQCD is located almost on these two curves.  
The two curves almost coincide up to CP and then deviate. 
In the large $\mu$ region, the curve of $T_{\rm max}$ lies below that of $T_{\chi, \rm max}$.  
This fact supports our assumption that, at least up to CP, the $T_{\rm max} (\mu )$ curve represents the limiting temperature of the baryon gas model. 

We also show the temperature $T_{\rm max, L}$ where the $T^5$-normalized trace anomaly $\Delta/T^5$ (not $(\Delta -\mu n_{\rm B})/T^5$) obtained by using the average value of $\Delta/T^4$ of the LQCD data in Ref.~\cite{Borsanyi:2012cr}. 
Since the lattice data is discretized in $T$, we regard the temperature interval as the systematic error bar. 
The $T_{\rm max,L}$ is located just above the crossover line and is comparable to $T_{\rm max}$ in the HRG model with EVE, but is somewhat smaller at $\mu =0$. This fact may indicate that, in the crossover transition, the partial transition from hadron to quark matter takes place below the limiting temperature of the baryon. 
In our HRG model with EVE, the absolute value of $\mu n_{\rm B}/T^5$ is about $20$ percent of that of $\Delta/T^5$ of the LQCD calculation when $\mu =0.4$~GeV and $T\sim 0.170$ GeV. 
Since $n_{\rm B}$ increases as $T$ increases in this region, the maximum temperature of $(\Delta -\mu n_{\rm B})/T^5$ will be somewhat smaller than that of $\Delta/T^5$.

\begin{figure}[t]
\centering
\centerline{\includegraphics[width=0.40\textwidth]{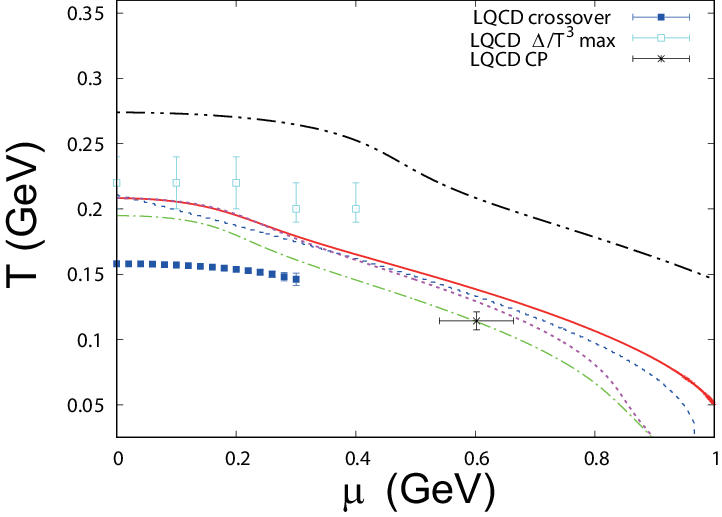}}
\caption{The dash-dot-dotted line shows the $T_{\rm zero}^\prime $ where $\Delta_{\rm B}$ vanishes in the HRG model with EVE.   
The solid (dotted )line shows the temperature $T_{\rm max}^{\prime}$ ($T_{\rm max}^{\prime\prime}$) where $\beta^3\Delta_{\rm B}$ has its maximum for the fixed value of $\gamma$ ($\mu$). 
The dashed line represents the curve determined by the condition $n_{\rm b}=1/(2v_{\rm B}$).  
The open squares with the error bar show the temperature $T_{\rm max,L}^{\prime\prime}$ where $\Delta/T^3$ has its maximum for the fixed value of $\mu$ obtained by using the LQCD data in Ref. ~\cite{Borsanyi:2012cr}.  
The meaning of the other symbols and the dash-dotted line are the same as in Fig.~\ref{Fig_max_min}.  
 }
 \label{Fig_max_min_gamma}
\end{figure}

Figure~\ref{Fig_max_min_gamma} shows the $\mu$-dependence of the temperature $T_{\rm zero}^\prime$ where $\Delta_{\rm B}$ vanishes in the HRG model with EVE. 
The $T_{\rm zero}^\prime$ is the limiting temperature of the HRG model with EVE obtained by the c-theorem like condition (\ref{effg_c_beta_1}). 
As in the case of Fig.~\ref{Fig_max_min}, at small $\mu$, the curve of $T_{\rm zero}^\prime$ is much higher than the LQCD crossover line.  
At $\mu =\mu_{\rm CP}$, $T_{\rm zeo}^\prime$ is much higher than the temperature $T_{\rm CP}$, where $\mu_{\rm CP}$ and $T_{\rm CP}$ are the baryon number chemical potential and the temperature at the critical point predicted by LQCD~\cite{Shah:2024img}, respectively.   

In Fig.~\ref{Fig_max_min_gamma}, 
the curves of $T_{\rm max}^\prime$ where $\beta^3\Delta_{\rm B}$ have its maximum for the fixed value of $\gamma$ in the HRG model with EVE are also shown. 
The $T_{\rm max}^\prime$ is the limiting temperature of the HRG model with EVE obtained by the strong condition (\ref{effg_c_beta_2}). 
The curve of $T_{\rm max}^\prime$ is somewhat higher than $T_{\rm max}$ but is close to the curve determined by the condition $n_{\rm b}=1/(2v_{\rm B})$. 
Note that, at $\mu =0$, the condition $n_{\rm b}=n_{\rm a}=1/(2v_{\rm B})$ is satisfied at $T=T_{\rm RWL}=210.3$~GeV~\cite{Oshima:2026bub}. 
This indicates that the condition $n_{\rm b}\le 1/(2v_{\rm B})$ gives the weak limitation condition for the HRG model with EVE.  

We also show the temperature $T_{\rm max, L}^{\prime\prime}$ where the $T^3$-normalized trace anomaly $\Delta/T^3(=\beta^3\Delta )$ obtained by using the average value of $\Delta/T^4$ of the LQCD data in Ref.~\cite{Borsanyi:2012cr}. 
Since the lattice data is discretized in $T$, we regard the temperature interval as the systematic error bar.  
The $T_{\rm max,L}^\prime$ is located just above $T_{\rm max}^\prime$ in the HRG model with EVE. 
Note that $T_{\rm max,L}^{\prime\prime}$ is obtained for the fixed value of $\mu$, not for $\gamma$. 
Hence, we also show $T_{\rm max}^{\prime\prime}$ where $\beta^3\Delta$ has its maximum for the fixed value of $\mu$ in the HRG model with EVE. 
The $T_{\rm max,L}^{\prime\prime}$ is also somewhat larger than $T_{\rm max}^{\prime\prime}$.

\section{Excluded volume effects in the meson gas} 
\label{EMESON}

In Fig.~\ref{Fig_cmu_mu=0}, it seems that $g_\mu (T)$ in the HRG model with EVE is consistent with the LQCD result up to $T\sim 0.2$ GeV where $\Delta/T^4$ has its maximum. 
However, this consistency may be accidental since it depends on the meson gas model. 
The interaction between mesons may also be important. 
Here, we assume here that the pressure of the meson gas is given by a similar form as that of the baryon gas with EVE;
\begin{eqnarray}
P_{\rm M}(T,\mu )&=&{T\over{v_{\rm M}}}\log{[1+v_{\rm M}M(T)]}, 
\label{Pb_EVE_Boltz_M}
\\
M(T)&=&\sum_i M_i(T), 
\label{MT}
\end{eqnarray}
where
\begin{eqnarray}
M_i(T)&=&{g_{{\rm M},i}\over{2\pi^2}}\int_0^\infty dp p^2e^{-\sqrt{p^2+m_{{\rm M},i}^2}/T}. 
 \label{MTi}
\end{eqnarray}
Here $m_{{\rm M},i}$, $g_{{\rm M},i}$ and $v_{\rm M}$ are the mass and the spin degeneracy of the $i$-th meson, and the volume of a meson, respectively. 
We call this model the HRG model with EVE2. 
Note that in the HRG model with EVE2, the EVE for baryons is also included.

\begin{figure}[t]
\centering
\centerline{\includegraphics[width=0.40\textwidth]{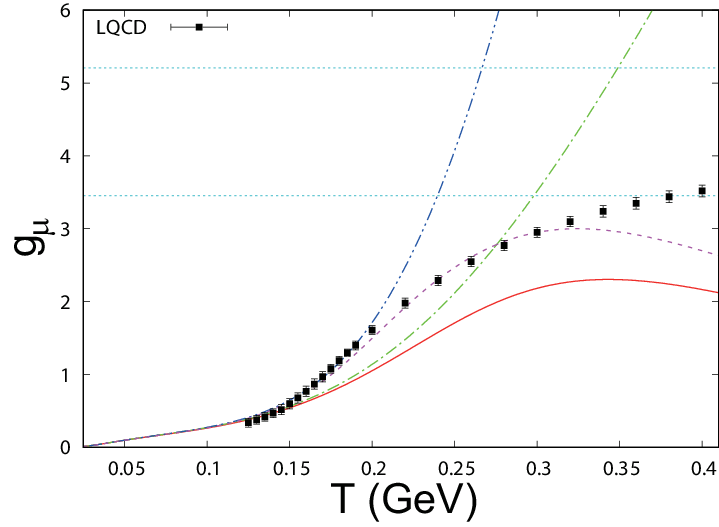}}
\caption{The $T$-dependence of the effective degrees of freedom $g_{\mu}$ at $\mu =0$.  
The dashed and dash-dot-dotted lines show the results in the HRG model with EVE2 and without EVE2, respectively. 
The solid and dash-dotted lines show the meson contribution $g_{\mu, M}$ in the HRG model with and without EVE2, respectively. 
We set $r_{\rm M}=0.3$~fm in the HRG model with EVE2.  
The upper (lower) dotted line shows result in the ideal massless three flavor QGP with (without) the gluon contributions. 
The squares with error bar show the LQCD results in Ref.~\cite{Borsanyi:2012cr}. 
 }
 \label{Fig_cmu_mu=0_M_03}
\end{figure}

Figure~\ref{Fig_cmu_mu=0_M_03} shows $g_\mu (T)$ as a function of $T$ in the HRG model with and without EVE2 when $\mu =0$. 
We set $v_{\rm M}={4\pi R_{\rm M}^3\over{3}}$ with $r_{\rm M}=0.3$ fm.  
It is seen that $g_\mu$ in the HRG with EVE2 is in good agreement with the LQCD result up to $T\sim 0.3$ GeV. 
However, this modification does not change $\chi_2^{\rm B}$ and the obtained $\chi_2^{\rm B}$ is still smaller than the LQCD result when $T>T_{\chi, \rm max}$=0.195 GeV.   
To reproduce the LQCD result of $\chi_2^{\rm B}/T^2$, the deconfined quark contribution needs to be added to the HRG model with EVE2. 
Once the quark contribution is added, the EDOF of the HRG model overshoots the LQCD result.  
Hence, the meson contribution should be suppressed more strongly.  

\begin{figure}[t]
\centering
\centerline{\includegraphics[width=0.40\textwidth]{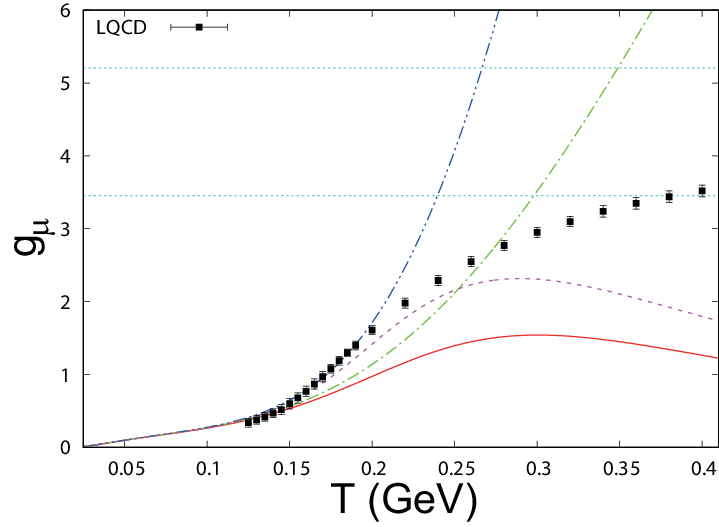}}
\caption{The $T$-dependence of the effective degrees of freedom $g_{\mu}$ at $\mu =0$.  
The dashed, and dash-dot-dotted lines show the results in the HRG model with and without EVE2, respectively. 
The solid, and dash-dotted lines show the meson contribution $g_{\mu, M}$ in the HRG model with and without EVE2, respectively. 
We set $r_{\rm M}=0.4$~fm in the HRG model with EVE2.  
The upper (lower) dotted line shows result in the ideal massless three flavor QGP with (without) the gluon contributions. 
The squares with error bar show the LQCD results in Ref.~\cite{Borsanyi:2012cr}. 
 }
 \label{Fig_cmu_mu=0_M_04}
\end{figure}

Figure~\ref{Fig_cmu_mu=0_M_04} is the same as Figure~\ref{Fig_cmu_mu=0_M_04} but $r_{\rm M}=0.4$ fm is used. 
It is seen that $g_\mu$ in the HRG with EVE2 is somewhat smaller than the LQCD result in the high temperature region. 
We assume that this difference will be compensated for by the contribution of quarks and gluons. 

\begin{figure}[t]
\centering
\centerline{\includegraphics[width=0.40\textwidth]{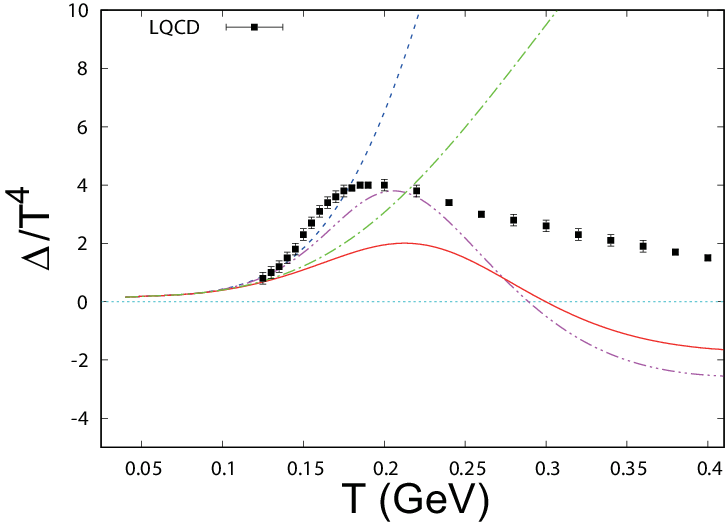}}
\caption{The $T$-dependence of the normalized trace anomaly $\Delta/T^4$ at $\mu =0$.  
The dashed and dash-dot-dotted lines show the results in the HRG model with EVE2 and without EVE2, respectively. 
The solid and dash-dotted lines show the meson contribution $\Delta_{\rm M} /T^4$ in the HRG model with and without EVE2, respectively. 
We set $r_{\rm M}=0.4$~fm in the HRG model with EVE2.  
The dotted line shows result in the ideal massless three flavor QGP with (without) the gluon contributions. 
The squares with error bar show the LQCD results in Ref.~\cite{Borsanyi:2012cr}. 
 }
 \label{Fig_ta_mu=0_M_04}
\end{figure}

Figure~\ref{Fig_ta_mu=0_M_04} shows $\Delta /T^4$ as a function of $T$ in the HRG model with and without EVE2 when $\mu =0$ and $r_{\rm M}=0.4$ fm.  
The result of the HRG with EVE2 is somewhat smaller than the LQCD data. 
Again, we assume that this difference will be compensated for by the contribution of quarks and gluons. 

\begin{figure}[t]
\centering
\centerline{\includegraphics[width=0.40\textwidth]{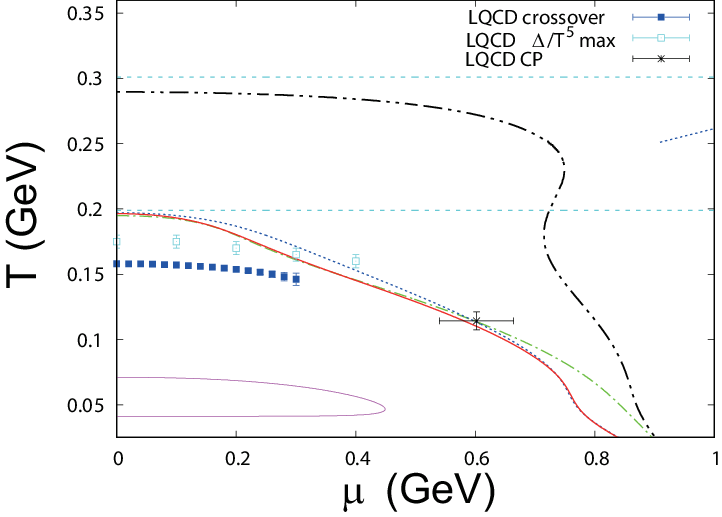}}
\caption{The dash-dot-dotted (upper dashed) line shows the $T_{\rm zero2}$ ($T_{\rm zero2M}$) where $\Delta-\mu n_{\rm B}$ ($\Delta_M$) vanishes in the HRG model with EVE2. 
The dotted  (lower dashed) line shows the temperature $T_{\rm max2}$ ($T_{\rm max2M}$) where $(\Delta-\mu n_{\rm B})/T^5$ ($\Delta_{\rm M}/T^5$) has its maximum for the fixed value of $\mu$. 
The thin solid line in the low $T$ and the small $\mu$ region is the local maximum or minimum of $(\Delta-\mu n_{\rm B})/T^5$. 
The meaning of the solid and dash-dotted lines and the symbols are the same as in Fig.~\ref{Fig_max_min}. 
 }
 \label{Fig_max_min_M}
\end{figure}

Figure~\ref{Fig_max_min_M} shows the $\mu$-dependence of the temperature $T_{\rm zero2}$ where $(\Delta-\mu n_{\rm B})$ vanishes in the HRG model with EVE2. 
The $T_{\rm zero2}$ is the limiting temperature of the HRG model with EVE obtained by the c-theorem like condition (\ref{effg_c_T_1}).  
At small $\mu$, the curve of $T_{\rm zero2}$ is much higher than the LQCD crossover line and, at $\mu =0$,  is close to the transition temperature $T_{\rm d}\sim 0.285$ GeV in the pure gluonic theory.~\cite{Borsanyi:2022xml,Giusti:2025fxu,Fujimoto:2025sxx}.  
At $\mu =\mu_{\rm CP}$, $T_{\rm zero2}$ is much higher than $T_{\rm CP}$. 
The temperature $T_{\rm zero2M}$ ($T_{\rm max2M}$) where $\Delta_{\rm M}$ vanishes (has its maximum) is constant and is 0.301 (0.199) GeV, since $\Delta_{\rm M}$ does not depend on $\mu$.

In Fig.~\ref{Fig_max_min_M}, the $\mu$-dependence of $T_{\rm max2}$ where $(\Delta -\mu n_{\rm B})/T^5$ has its maximum for the fixed value of $\mu$ in the HRG model with EVE2 is also shown. 
The $T_{\rm max2}$ is the limiting temperature of the HRG model with EVE obtained by the strong condition (\ref{effg_c_T_2}). 
We see that the critical point predicted by LQCD is located almost on the curve of $T_{\rm max2}$.  
The curve of $T_{\rm max2}$ is close to that of $T_{\rm max}$ and crosses the curve of $T_{\chi,\rm max}$ in the vicinity of the CP ($(\mu_{\rm CP}, T_{\rm CP})=(0.6021\pm 0.0621{\rm GeV},0.1143  \pm 0.0069 {\rm GeV})$). 
The cross point is given by
\begin{eqnarray}
(\mu_{\rm CRP},T_{\rm CRP})=(0.594~{\rm GeV}, 0.115~{\rm GeV}), 
 \label{Eq_cross point}
\end{eqnarray}
and is very close to the CP. 
This indicates that the limiting temperature of the HRG model with EVE contains some information about the critical point, although the model has no mechanism for the chiral phase transition. 

There is a isolated curve of $T_{\rm max2}$ in the high $T$ and $\mu$ region. 
However, this peak structure is beyond the curve of $T_{\rm zero2}$ and is not important for our discussions. 
In the low $T$ and the small $\mu$ region, there is also the curve where $(\Delta-\mu n_{\rm B})/T^5$ has its local maximum or minimum. 
The upper part of this curve is the local minimum line, and the lower part is the local maximum line.  
As was already discussed in Sec.~\ref{HRGEDOF}, this structure is induced by the pseudo Nambu-Goldstone boson, namely, the pion contribution. 
This curve may divide the low temperature and density phase into three parts. 

\begin{figure}[t]
\centering
\centerline{\includegraphics[width=0.40\textwidth]{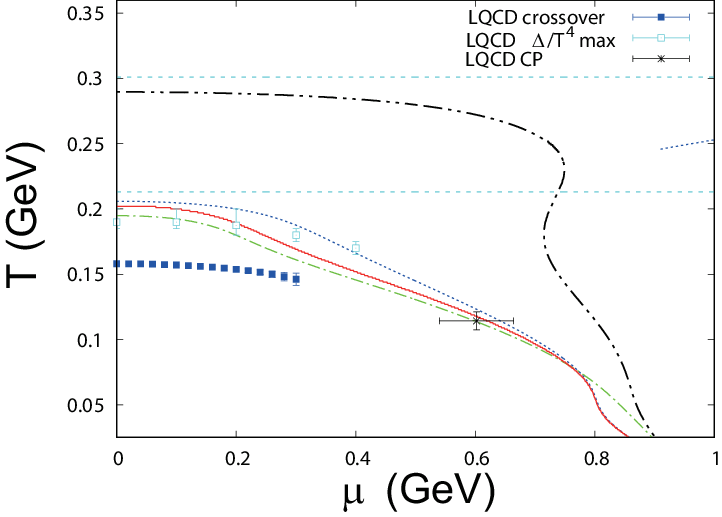}}
\caption{The dash-dot-dotted (upper dashed) line shows the $T_{\rm zero2}$ ($T_{\rm zeroM}$) where $\Delta-\mu n_{\rm B}$ ($\Delta_M$) vanishes in the HRG model with EVE2. 
The dotted  (solid, lower dashed) line shows the temperature $T_{\rm max2t}$ ($T_{\rm maxt}$,$T_{\rm max2Mt}$) where $(\Delta-\mu n_{\rm B})/T^4$ (($\Delta_{\rm B}-\mu n_{\rm B})/T^4$,$\Delta_{\rm M}/T^4$) has its maximum for the fixed value of $\mu$. 
The open squares with the error bar show the temperature $T_{\rm max,Lt}$ where $\Delta/T^4$ has its maximum obtained by using the LQCD data in Ref. ~\cite{Borsanyi:2012cr}. 
The meaning of the dash-dotted line and the other symbols are the same as in Fig.~\ref{Fig_max_min}. 
 }
 \label{Fig_max_min_M_t}
\end{figure}

However, it should be remarked that this local peak structure induced by pion disappears if we use a dimensionless scale parameter $t=\log{(T/T_0)}$ where $T_0$ is a constant parameter with the same dimension as $T$. Such a dimensionless scale parameter is usually used in the renormalization group equation. 
In this case, the evolution equation is given by 
\begin{eqnarray}
{\partial g_\mu (T) \over{\partial t}}=T{\partial g_\mu (T) \over{\partial T}}&=&{\Delta -\mu n_{\rm B}\over{T^4}}
\label{effg_c_small_t_1}
\end{eqnarray}
Note that this modification does not change $T_{\rm zero2}$ where the right hand side of the equation vanishes. 
However, the temperature where the right hand side has its (local) maximum changes or disappears. 
Figure~\ref{Fig_max_min_M_t} is the same as Fig.~\ref{Fig_max_min_M} but the evolution equation (\ref{effg_c_small_t_1}) is used. 
We see that the local peak structure induced by pion disappears.
On the other hand, qualitative characteristics of the the global maximum structure is retained although $T_{\rm max2t}$ is somewhat larger than $T_{\rm max2}$ in the small $\mu$ region. 
It should be also remarked that $T_{\rm max2t}$ is close to $T_{\rm max2,Lt}$ obtained by using the LQCD result~\cite{Borsanyi:2012cr}. 
In this meaning, the $(t,\mu )$ coordinate system may be more preferable than $(T,\mu )$. 

\section{Summary}
\label{summary}

In summary, to investigate the limiting temperature of the hadron resonance gas model with EVE, we have studied the relation between the effective degrees of freedom and the trace anomaly in the HRG model. 
If we regard the thermodynamic relation as the evolution equation and define the effective degree of freedom as $P/T^4$ (or $P/\mu^4$), we obtain the equations that relate to the trace anomaly. 
The structures of the equations resemble the equation of the so-called c-theorem in the two dimensional conformal field theory. 
There is a stationary point where the (modified) trace anomaly vanishes and the scale invariance is restored. 

We consider two types of the c-theorem like condition for the effective degrees of freedom of the HRG model with EVE.  
The first condition requires that the EDOF should not decrease as the energy scale increases. 
If we use the $(T,\mu)$ coordinate, 
according to this condition, the modified trace anomaly should not be negative. 
However, the limiting temperature obtained by this requirement is much higher than the crossover transition temperature predicted by the LQCD calculation ~\cite{Borsanyi:2020fev}, and, at $\mu =0$, it is close to the transition temperature $T_{\rm d}\sim 0.285$ GeV in the pure gluonic theory~\cite{Borsanyi:2022xml,Giusti:2025fxu,Fujimoto:2025sxx}.  
If we use the $(\beta,\gamma )$ coordinate, $T_{\rm zero}^\prime$ where $\Delta_{\rm B}$ vanishes at $\gamma =0$ is also close to $T_{\rm d}$. 

We also consider the stronger condition that the second derivative of the EDOF with respect to the scale parameter should not be negative. 
This condition is equivalent to requiring that the EDOF is convex downwards as a function of the scale parameter in the HRG model. 

If we use the $(T,\mu )$ coordinate, 
according to this condition, the $T^5$-normalized trace anomaly should not decrease as $T$ increases. 
The temperature at which the modified trace anomaly reaches its maximum is the limiting temperature of the HRG model with EVE. 
The obtained limiting temperature of the HRG model with EVE (EVE2) is consistent with the one obtained by using the normalized baryon fluctuation in the previous study and the LQCD predicted critical point~\cite{Shah:2024img}. 
This result supports our assumption that the temperature of the maximum $\chi_2^{\rm B}$ is the limiting temperature of the HRG model with EVE.  

If we use the $(\beta,\gamma )$ coordinate, $T_{\rm max}^\prime$ where $\beta^3\Delta_{\rm B}$ has its maximum for the fixed value of $\gamma$ is somewhat higher than $T_{\rm max}$ obtained by using the $(T,\mu )$ coordinate but is close to the curve determined by the condition $n_{\rm b}=1/(2v_{\rm B})$. 
In particular, at $\mu =0$, $T_{\rm max}^\prime$ almost coincides with the Roberge-Weiss like temperature $T_{\rm RWL}$~\cite{Oshima:2026bub} which is the limiting temperature of the HRG model with EVE at $\mu =i(2k+1)\pi T$ where $k$ is any integer. 
This indicates that the condition (\ref{effg_c_beta_2}) is weaker than (\ref{effg_c_T_2}). 
There is an open question which coordinate system is most appropriate to determine the limiting temperature. 
A more fundamental and microscopic understanding of the increase and decrease in degrees of freedom is needed. 

It should be also remarked that these limiting temperatures are for the HRG with EVE but not for QCD. 
In fact, at $\mu =0$, the EDOF is convex upwards in the high temperature region. 
Hence, hadron matter is expected to transition to quark matter or exotic new matter in the high temperature region. 
Furthermore, since our transition scenario is based on the observation on the crossover transition at $\mu =0$, it is an open question whether these limiting conditions are satisfied or not, beyond the critical point where the fluctuations shows the divergent behavior.  

There is also an open question why the temperature of the maximum of the $T^5$-scaled modified trace anomaly is so close to that of the normalized baryon number fluctuation. 
It is important that these temperatures are just below the temperature determined by the condition $n_{\rm b}=1/(2v_{\rm B})$. 
This indicates that the suppression factor $1/(1+v_{\rm B}n_{\rm b})$ plays a crucial role. 
The condition $n_{\rm b}=1/(2v_{\rm B})$ is also related to the $R=0$ criterion in the thermodynamic geometry~\cite{Oshima:2026bub}. 
Detailed study of the role of the suppression factor is needed in the future.

It is very interesting that, in the HRG model with EVE2, the curve of the maximum $\Delta -\mu n_{\rm B}$ crosses with the curve of the maximum $\chi_2^{\rm B}$ near the CP predicted by the LQCD calculation~\cite{Shah:2024img}. 
This indicates that the limitation of the HRG model with EVE has some information about the critical point, although the model has no mechanism for the chiral phase transition. 
It was pointed out that, near the critical point, the ordering density is a linear combination of the scalar density, the net baryon number density, and the energy density rather than the pure scalar density itself~\cite{Fujii:2004jt}.  
The effects of the baryon number density and the trace anomaly may be more important in the large $\mu$ region. 
Further study to investigate the nature of the transition beyond CP is needed.


\begin{acknowledgments}
The authors are deeply grateful to Motoi Tachibana for helpful and important discussions.
H.K. also thanks Hajime Aoki for useful discussions on conformal field theory. 
This work is supported in part by Grants-in-Aid for Scientific Research from the Japan Society for the Promotion of Science (JSPS) KAKENHI (Grant No. JP22H05112).
\end{acknowledgments}


\bibliography{ref.bib}

\end{document}